\documentclass[12pt,a4paper]{article}

\usepackage{amssymb}
\usepackage{physics}

\usepackage{authblk}

\usepackage{hyperref}

\DeclareMathOperator{\Span}{span}
\DeclareMathOperator{\id}{id}
\DeclareMathOperator{\diag}{diag}
\DeclareMathOperator{\Div}{div}
\DeclareMathOperator{\Curl}{curl}

\DeclareMathOperator{\dal}{\Box}
\DeclareMathOperator{\hodge}{\star}
\newcommand{\codd}{\delta}

\newcommand{\lie}[1][Z]{
	\mathcal{L}
	_{#1}}
\newcommand{\inn}[1][Z]{i_{#1}}
\newcommand{\est}[1][\dd{z}]{\varepsilon_{#1}}

\newcommand{\Puno}[1][]{\mathcal{P}_{#1}}
\newcommand{\Pdue}[1][]{\mathcal{Q}_{#1}}

\title{
Electromagnetism: an intrinsic approach to Hadamard's 
	method of descent}

\begin{document}

\author[1]{Giuliano Angelone}
\author[2,3]{Elisa Ercolessi}
\author[4,5]{Paolo Facchi}
\author[4,5]{Rocco Maggi}
\author[6,7]{Giuseppe Marmo}
\author[4,5]{Saverio Pascazio}
\author[4,5]{Francesco V. Pepe}

\affil[1]{Dipartimento di Matematica, Sapienza
Universit\`{a} di Roma, 00185 Rome,
Italy}

\affil[2]{Dipartimento di Fisica e Astronomia, Universit\`{a} di Bologna, 40127 Bologna, Italy}

\affil[3]{INFN, Sezione di Bologna, 40127 Bologna, Italy}

\affil[4]{Dipartimento di Fisica, Universit\`{a} di Bari, 70126 Bari, Italy}
\affil[5]{INFN, Sezione di Bari, 70126 Bari, Italy}
\affil[6]{Dipartimento di Fisica, Universit\`{a} di Napoli Federico II, 80126 Naples, Italy}
\affil[7]{INFN, Sezione di Napoli, 80126 Naples, Italy}

	\date{\today}

	\maketitle

\begin{abstract}

We present a systematic geometric framework for the dimensional reduction of classical electromagnetism based on the concept of descent along  vector fields of invariance. 
By exploring the interplay between the Lie derivative and the Hodge star operator, we  implement descent conditions on differential forms that reduce Maxwell's equations in four-dimensional spacetime  to electromagnetic theories in lower dimensions. We also consider multiple descent along pairwise commuting  vector fields of invariance, yielding a finer decomposition of Maxwell's equations. Our results provide a unified and geometrically transparent interpretation of dimensional reduction, with potential applications to field theories in lower-dimensional spacetimes.
\end{abstract}

\section{Introduction}
The method of descent was used by Hadamard to discuss partial differential equations with a different number of spatial dimensions~\cite{hada}. As a paramount example, the solutions of the cylindrical wave equation, and of its Cauchy problems, are related to those of the spherical wave equation. In more prosaic terms, the scalar wave equation in two spatial dimensions is regarded, and solved, as an instance of the scalar wave equation in three spatial dimensions.

In some recent papers~\cite{descent,descent2,descent3}, this idea has been used to perform the dimensional reduction of electromagnetism, and of the Dirac equation, in Minkowski spacetimes. The low-dimensional theories obtained by descent are conceived as describing the special cases of the higher-dimensional ones, in which all the relevant quantities are uniform along the additional spatial directions.
Remarkably, when applied to the equations governing a vector or a spinor field, this procedure may provide novel sets of decoupled low-dimensional equations, describing independent evolutions of distinct sets of physical quantities. This phenomenon has no analogue in the scalar case. 

In the physics literature, seminal ideas concerning electromagnetism were proposed and discussed by Ehrenfest~\cite{ehrenfest}. The problem has been considered by a number of authors~\cite{lapidus,moreno,wheelerem}, also recently~\cite{mcdonald,boito,GBB,BBG}.

By using  Cartesian coordinates, Hadamard's ``descent procedure'' has been applied ``componentwise'' to Maxwell equations in Ref.~\cite{descent}. As a result, the components of all electromagnetic quantities are partitioned into independently evolving low-dimensional sectors. However, the prominent role played by the coordinate system does not allow us to foresee what would happen when using generalized coordinates, or how to generalize such procedure to non-Abelian gauge theories, such as Yang--Mills, or to Einstein equations.

What we have learned from electromagnetism is that the starting differential equation in a $(3+1)$-dimensional Minkowski spacetime separates into  equations which are independent from each other. If we think of the procedure in the opposite direction, it suggests a way to put together, that is, to unify low-dimensional equations which appear to be distinct and to have different interpretations.

As a preliminary investigation, aimed at better understanding how the descent procedure should be applied to ``vector-valued'' fields, in order to extend it to more general theories, in this work we will reconsider the case of electromagnetism (in vacuum and in a Minkowski spacetime) with a coordinate-free approach.

The article is structured as follows. 
In Sec.~\ref{sec:recap} we review the method of Hadamard's descent, and show how its basic idea can be applied to Maxwell equations in Cartesian coordinates to perform a dimensional reduction of electromagnetism.
In Sec.~\ref{sec:intrinsic} we discuss our spacetime settings, relating our basic assumption to the wave operator, and formulate electromagnetism in terms of differential forms. Our main contributions are in Secs.~\ref{sec:reformulation}  and~\ref{sec:seconddescent}.  
In Sec.~\ref {sec:reformulation}, Hadamard's ideology, originally applied to scalar functions, is first reinterpreted for differential forms, and then applied to electromagnetism. This results in a coordinate-free description of the aforementioned separation when we descend to $(2+1)$. In particular, the resulting equations can be interpreted as pullbacks of equations describing distinct electromagnetic models in a $(2+1)$-dimensional spacetime.
In Sec.~\ref{sec:seconddescent} we then reiterate our analysis by considering the descent to $(1+1)$.

\section{The method of descent}\label{sec:recap}
In this section we will first recall the original meaning carried by the expression \emph{method of descent} in Hadamard's work. To this end, we will review its application to the scalar wave equation, focusing on some results that are particularly relevant for physics. Then we will show how the underlying idea has been borrowed in Ref.~\cite{descent} to perform a dimensional reduction of electromagnetism in Cartesian coordinates, discussing the main results.

\subsection{Hadamard's method of descent and the scalar wave equation}\label{sec:HDMwaveeq}
In his lectures on partial differential equations~\cite{hada}, Hadamard considers the homogeneous wave equation with different numbers of spatial dimensions. 
The 3-dimensional or spherical wave equation, written in Cartesian orthonormal coordinates as
\begin{align}\label{eq:WaveHom3D}
	\frac{1}{c^2}\frac{\partial^2 u}{\partial t^2}
	-\frac{\partial^2 u}{\partial x^2} -
	\frac{\partial^2 u}{\partial y^2} -
	\frac{\partial^2 u}{\partial z^2}=0\,,
\end{align}
is first introduced as describing the propagation of sound in ordinary space, filled by a homogeneous gas. Then the 2-dimensional or cylindrical wave equation,
\begin{equation}\label{eq:WaveHom2D}
	\frac{1}{c^2}\frac{\partial^2 v}{\partial t^2}
	-\frac{\partial^2 v}{\partial x^2}
	-\frac{\partial^2 v}{\partial y^2}
	=0\,,
\end{equation}
is introduced as a special case of the spherical equation, describing an acoustic wave under the assumption of uniformity of all the relevant physical quantities along the $z$ direction. 
First of all, the solutions of Eq.~\eqref{eq:WaveHom2D} are all and only the solutions of Eq.~\eqref{eq:WaveHom3D} that are independent of the $z$ coordinate. These considerations extend to Cauchy problems: namely, $v=v(t,x,y)$ is a solution of the cylindrical problem with (sufficiently regular) initial data
\begin{align}\label{eq:Cauchy2D}
	v(0,x,y)=v_0(x,y)\,,
	\qquad
	\frac{\partial v}{\partial t}(0,x,y)=v_1(x,y)\,,
\end{align}
if and only if $u(t,x,y,z)=v(t,x,y)$ is a $z$-independent solution of the spherical problem with the same initial data,
\begin{align}\label{eq:Cauchy3D}
	u(0,x,y,z)=v_0(x,y)\,,
	\qquad
	\frac{\partial u}{\partial t}(0,x,y,z)=v_1(x,y)\,.
\end{align}
The condition is obviously sufficient. On the other hand, if $u=u(t,x,y,z)$ is a solution of the above problem, then $\tilde u(t,x,y,z)=u(t,x,y,z+h)$ is in turn a solution. Therefore, due to uniqueness of the classical solution of the wave equation in arbitrary spatial dimensions, the condition is also necessary.

Hadamard refers to this framework as \emph{a method of descent}, and sums it up in the motto \emph{he who can do more can do less}. The idea is that we can always solve a differential problem, when it can be regarded as a special case of a more general problem, which we are already able to solve, and whose solutions are independent of the additional variables. Hadamard consistently resorted to the method of descent to solve differential problems, as well as to investigate the properties of their solutions.

Finally, let us take into account the inhomogeneous problem: the solutions of the cylindrical inhomogeneous equation
\begin{align}\label{eq:cylindricalNH}
	\frac{1}{c^2}\frac{\partial^2 u}{\partial t^2}
	-\frac{\partial^2 u}{\partial x^2} 
	-\frac{\partial^2 u}{\partial y^2} 
	=s(t,x,y)\,,
\end{align}
are all and only the  solutions of the spherical inhomogeneous equation\begin{align}\label{eq:sphericalNH}
	\frac{1}{c^2}\frac{\partial^2 u}{\partial t^2}
	-\frac{\partial^2 u}{\partial x^2} 
	-\frac{\partial^2 u}{\partial y^2} 
	-\frac{\partial^2 u}{\partial z^2}
	=s(t,x,y)\,,
\end{align}
that are independent of $z$, and these considerations extend to Cauchy problems, again by uniqueness of the classical solution. 

\subsection{Dimensional reduction of electromagnetism by 
	descent}\label{sec:CartesianReduction}
The fundamental idea behind Hadamard's method of descent can be explored as a way to perform a dimensional reduction of electromagnetism. 
This problem has been studied in Ref.~\cite{descent} for electrodynamics in vacuum, with external sources, in a Minkowski spacetime.
The low-dimensional models describe all the possible realizations of electrodynamics in $(3+1)$ that are characterized by translational invariance along either one or two independent spatial directions, in some inertial reference frame. This procedure yields two models for $(2+1)$ and three, one of which trivial (in absence of monopoles) for $(1+1)$. Each such model is both a low-dimensional relativistic theory, and a special case of ordinary electromagnetism, reproducing only some of its traits. Different models are formulated in terms of distinct physical quantities, they can live side by side, but cannot talk to each other, just like electrostatics and magnetostatics.

In each lower dimension, some of the usual features are inherited by all models, others are split among them, still others are lost. In particular, only one model, in each lower dimension, allows for electric charge and charge imbalance. These models are special, as they retain the mathematical structure of an Abelian gauge theory, and can be regarded as ``scaled down'' versions of ordinary electromagnetism. 
For this reason, they have usually been considered in the literature as \emph{the} low-dimensional electromagnetic theories. The descent procedure allows us to recover these theories, along with other, no less valid (although lesser known) theories, with strikingly different properties.

\subsubsection{Descent to \texorpdfstring{$(2+1)$}{(2+1)}} \label{sec:firstdesc}

We now proceed to analyze the dimensional reduction from $(3+1)$ to $(2+1)$, following the analysis in Ref.~\cite{descent}
In an inertial reference frame, electrodynamics in vacuum with external sources can be formulated in terms of the electric field $\vb*{E}$, and the magnetic field $\vb*{B}$, which are governed by Maxwell equations (henceforth written in natural and rationalized units),
\begin{align}\label{eq:EBhomo}
	\Div \vb*{B}=0\,, 
	\qquad 
	\Curl\vb*{E}+ \frac{\partial \vb*{B}}{\partial t} =\vb*{0}\,,
	\\\label{eq:EBsource}
	\Div \vb*{E}=\rho\,, 
	\qquad 
	\Curl \vb*{B}-\frac{\partial \vb*{E} }{\partial t}
	=\vb*{j}\,.
\end{align}
Consistency of~\eqref{eq:EBsource} implies that the charge density $\rho$ and the current density $\vb*{j}$ must satisfy the continuity equation
\begin{align}\label{eq:continuity}
	\frac{\partial \rho}{\partial t} +\Div \vb*{j}=0\,,
\end{align}
expressing the conservation of the electric charge. 

Considered a Cartesian coordinate system, the descent to $(2+1)$ is performed by requiring that all electromagnetic quantities be independent of the $z$ coordinate, that is, by imposing the (infinitesimal) ``descent conditions''
\begin{align}\label{eq:FirstDescCond}
	\partial_z E_i = 0\,,
	\qquad
	\partial_z B_i = 0\,,
	\qquad
	\partial_z \rho = 0\,,
	\qquad
	\partial_z j_i = 0\,,
\end{align}
with $i=x,y,z$.
Then Maxwell equations~\eqref{eq:EBhomo}--\eqref{eq:EBsource} split up into two uncoupled subsystems, each of four equations. One subsystem,
\begin{gather}\label{eq:EEB1}
	\partial_t B_z+\partial_x E_y-\partial_y E_x=0\,,\\
	\partial_x E_x+\partial_y E_y=\rho\,,\\
	-\partial_t E_x+\partial_yB_z=j_x\,,\\
	\label{eq:EEB4}
	-\partial_t E_y-\partial_x B_z=j_y\,,
\end{gather}
governs the dynamics of $E_x$, $E_y$, and $ B_z$, and has three inhomogeneous equations, with source terms $\rho$, $j_x$, and $j_y$, satisfying the reduced continuity equation $\partial_t \rho +\partial_x j_x+\partial_y j_y=0$. The other subsystem,
\begin{gather}\label{eq:BBE1}
	\partial_x B_x + \partial_y B_y = 0\,,\\
	\partial_t B_x + \partial_y E_z = 0\,,
	\\
	\partial_t B_y - \partial_x E_z = 0\,,
	\\\label{eq:BBE4}
	-\partial_t E_z + \partial_x B_y - \partial_y B_x = j_z\,,
\end{gather}
is in terms of $B_x$, $B_y$, and $E_z$, and has  just one inhomogeneous equation, with source term $j_z$. 

Observe that the Cartesian components of the 
electromagnetic quantities have been partitioned into two independent electromagnetic sectors, giving rise to two $(2+1)$-dimensional models. These sectors, and the corresponding models, can be labelled in terms of the number of electric and magnetic components involved. The scaled down version of ordinary electromagnetism is the EEB model. The BBE model can be formulated  in terms of a vector field, originating from a scalar potential, which can be coupled to a scalar charge. Both models allow for wave propagation: the spherical wave equation satisfied by each field component in $(3+1)$ is reduced to a cylindrical equation.

The above picture can be compared to stationary electromagnetism. When all electromagnetic quantities do not depend on time, two independent sets of stationary equations are obtained, yielding two separate stationary theories: electrostatics, in terms of the electric field produced by static charges, and magnetostatics, in terms of the magnetic field produced by stationary currents.

\subsubsection{Descent to \texorpdfstring{$(1+1)$}{(1+1)}} \label{sec:seconddesc}
Let us now perform the descent to $(1+1)$, by assuming that all electromagnetic quantities are independent of both $y$ and $z$, that is, by imposing the descent conditions along $z$ of Eq.~\eqref{eq:FirstDescCond}, together with the analogous conditions along $y$, 
\begin{align}\label{eq:SecondDescCond}
	\partial_y E_i = 0\,,
	\qquad
	\partial_y B_i = 0\,,
	\qquad
	\partial_y \rho = 0\,,
	\qquad
	\partial_y j_i = 0\,.
\end{align} 
This is equivalent to performing a second descent along $y$ on the sectors arising from a first descent along $z$. As a result, each sector splits up into two uncoupled subsystems, of two equations each.

The EEB sector decouples into a purely electric sector,
\begin{gather}
	\partial_x E_x = \rho\,,\label{eq:E1}\\
	-\partial_t E_x = j_x\,,\label{eq:E2}
\end{gather}
governing  $E_x$, via two inhomogeneous equations, with source terms $\rho$ and $j_x$, satisfying the reduced continuity equation $\partial_t \rho +\partial_x j_x=0$, and a mixed sector in $E_y$ and $B_z$, 
\begin{gather}\label{eq:EB1}
	\partial_t B_z + \partial_x E_y = 0\,,
	\\\label{eq:EB2}
	-\partial_t E_y - \partial_x B_z = j_y,
\end{gather}
with one inhomogeneous equation, with source term $j_y$. 

Likewise, the BBE equations split up into a mixed subsystem in $E_z$ and $B_y$,
\begin{gather}\label{eq:BE1}
	\partial_t B_y - \partial_x E_z = 0\,,
	\\\label{eq:BE2}
	-\partial_t E_z + \partial_x B_y = j_z\,,
\end{gather}
with one inhomogeneous equation, with source term $j_z$, and a purely magnetic sector, made up of two homogeneous equations in $B_x$, 
\begin{gather}\label{eq:B1}
	\partial_x B_x = 0\,,
	\\\label{eq:B2}
	\partial_t B_x = 0\,.
\end{gather}

Now the Cartesian components of all electromagnetic quantities are divided into four independent electromagnetic sectors. Observe that the mixed subsystems~\eqref{eq:EB1}--\eqref{eq:EB2} and~\eqref{eq:BE1}--\eqref{eq:BE2} are turned into each other by a $\pi/2$ rotation around the $x$ axis (the only surviving spatial direction, determining a cylindrical symmetry). 

Therefore, we obtain three $(1+1)$-dimensional models: i) The purely electric model is the scaled down version of electromagnetism in $(3+1)$, and its field is instantaneous in the sources; ii) The purely magnetic model is sourceless and non-dynamical, its only possible solutions being uniform and stationary; iii) The mixed model can be described in terms of a vector field, originating from a scalar potential, which can be coupled to a scalar charge. It has several features in common with electromagnetism in $(3 + 1)$, such as an equal number of electric and magnetic components, as well as of homogeneous and inhomogeneous equations. Above all, it is the only model allowing for a ``true'' wave propagation: although the spherical equation governing each field component is reduced to a plane wave, this information is obviously redundant for the pure models.

\section{Settings}\label{sec:intrinsic}
In this section we will set the stage for our study. 
In Sec.~\ref{sec:exterior}, our assumptions on spacetime will be put in relation to the d'Alembert differential operator (see Sec.~\ref{sec:prelim} for a brief recap on how the basic notions are introduced). 
In Sec.~\ref{sec:electromagnetism}, electromagnetism will be formulated in terms of differential forms.

\subsection{Spacetime}\label{sec:exterior}
We assume spacetime to be a real four-dimensional affine space $M^4$. The translation group (which can be identified with the vector space associated to the affine structure, regarded as an Abelian group) acts freely and transitively on spacetime. Let $X_\mu$, with $\mu\in\{0,1,2,3\}$, be the infinitesimal generators of such group. Each of them is a complete vector field on $M^4$, generating a one-parameter subgroup of the translation group, and they commute with each other,
\begin{equation}
	\comm{X_\mu}{X_\nu}=0\,.
\end{equation}
In particular, $(X_0,X_1,X_2,X_3)$ gives a global frame for the tangent bundle~$TM^4$.

Given the crucial role played by the wave equation in the electromagnetic theory, we introduce the d'Alembert operator
\begin{equation}\label{eq:dalembert}
	\dal
	=\lie[X_0]\lie[X_0]
	-\lie[X_1]\lie[X_1]-\lie[X_2]\lie[X_2]-\lie[X_3]\lie[X_3]\,,
\end{equation}
where $\lie[X]$ is the Lie derivative with respect to the vector field $X$. This hyperbolic homogeneous differential operator is invariant under the action of the translation group, which is expressed, at infinitesimal level, by $\comm{\dal}{X_\mu}=0$. 
In a sense, the affine nature of spacetime can be regarded as a consequence of the request that we can define a homogeneous hyperbolic differential operator, and that the spacetime manifold is endowed with a group of translations, which are among its symmetries.

Furthermore, the d'Alembert operator brings about a Lorentzian structure on $M^4$~\cite{vinogradov}. Indeed, its principal symbol defines the symmetric and non-degenerate (2,0)-tensor field
\begin{equation}\label{eq:MinkT}
	\tilde\eta
	=\sigma(\dal)
	= X_0\otimes X_0-X_1\otimes X_1-X_2\otimes X_2-X_3\otimes X_3
	=\eta^{\mu\nu}X_\mu\otimes X_\nu
	\,,
\end{equation}
which is precisely the contravariant form of the Lorentzian metric
\begin{equation}
	\eta
	= \alpha^0\otimes \alpha^0-\alpha^1\otimes \alpha^1-\alpha^2\otimes \alpha^2-\alpha^3\otimes \alpha^3
	=\eta_{\mu\nu}\alpha^\mu\otimes \alpha^\nu\,,
\end{equation}
with $(\alpha^0,\alpha^1,\alpha^2,\alpha^3)$ being the coframe of $(X_0,X_1,X_2,X_3)$, defined by $\alpha^\mu (X_\nu) =\delta^\mu_\nu$. 
Both tensor fields are preserved by a group of affine transformations isomorphic to the Poincar\'{e} group. Moreover, both the frame $(X_0,X_1,X_2,X_3)$ and the coframe $(\alpha^0,\alpha^1,\alpha^2,\alpha^3)$ are orthonormal by construction, as the matrices $(\eta^{\mu\nu}) = \bigl(\tilde\eta(\alpha^\mu,\alpha^\nu)\bigr)$,
and $(\eta_{\mu\nu}) = \bigl(\eta(X_\mu,X_\nu)\bigr)$, are both equal to $\diag(+1,-1,-1,-1)$.

We choose the metric volume form as the basis 4-form
\begin{equation}\label{eq:volume}
	\Omega_\eta  
	= \alpha^0\wedge\alpha^1\wedge \alpha^2\wedge\alpha^3\,.
\end{equation}
This choice determines the Hodge star operator
\begin{equation}\label{eq:hodgeM}
	\hodge \omega  =  i_{\omega^\sharp} \Omega_\eta\,,
\end{equation}
the codifferential
\begin{equation}
	\codd=\hodge \dd{} \hodge\,,
\end{equation}
and the Laplace--Beltrami operator $\dd\codd+\codd\dd$. Then the d'Alembert operator~\eqref{eq:dalembert} and the Laplace--Beltrami operator coincide on the exterior algebra, namely,\footnote{Observe that neither operator in the above equation can be regarded as ``more general'' than the other one. On the one hand, the Laplace--Beltrami operator requires a pseudo-Riemannian structure, and  applies only to differential forms. On the other, a differential operator in the form~\eqref{eq:dalembert} requires a parallelizable manifold, can be applied to any tensor field, and a metric tensor is encapsulated in its principal symbol.}
\begin{equation}
	\dal \omega= -(\dd \codd +\codd \dd)\omega\,,
	\label{eq:wavelaplace}
\end{equation}
for any differential form $\omega$. This result is shown in Appendix~\ref{sec:DaleVsLB}, under slightly more general hypotheses (on the manifold, and the differential operator).

Finally, the 1-forms $\alpha^\mu$ are closed, since
\begin{equation}\label{eq:closed}
	\dd{\alpha^\mu }(X_\nu ,X_\rho )
	=X_\nu  \qty(\alpha^\mu  (X_\rho ))
	-X_\rho  \qty(\alpha^\mu  (X_\nu))
	-\alpha^\mu (\comm{X_\nu }{X_\rho})=0\,.
\end{equation}
Since $M^4$ is contractible, we can choose a set of functions $x^\mu$ such that $\alpha^\mu = \dd{x^\mu}$. In this way we have constructed  orthonormal global coordinates $(x^\mu)=(t,x,y,z)$, yielding respectively the frames $(\partial/\partial x^\mu)=(X^\mu)$, for the tangent bundle $TM^4$, and $(\dd{x^\mu})=(\alpha^\mu)$, for the cotangent bundle $T^*M^4$.

\subsection{Electromagnetism}\label{sec:electromagnetism}
The intrinsic formulation of electromagnetism in $M^4$ is given in terms of two differential 2-forms, the Faraday form $F$ and the Amp\`{e}re form $G$, and of the charge-current 3-form $J$. These forms are governed by the Faraday equation
\begin{align}\label{eq:dF}
	\dd F=0\,,
\end{align}
the Amp\`{e}re equation
\begin{align}\label{eq:dG}
	\dd G=J\,,
\end{align}
and a constitutive equation relating $F$ and $G$, ensuring that the differential problem is properly set. 

Observe that the Amp\`{e}re equation requires that the charge-current 3-form be closed,
\begin{equation}\label{eq:dJ}
	\dd J=0\,,
\end{equation}
whereas the Faraday equation implies the existence of a potential 1-form $A$, such that $F=\dd A$~\cite{hehl03, wheeler}. 

For electromagnetism in vacuum, we assume the constitutive equation\footnote{
	Although we will not insist on this distinction, it should be pointed out that while $F$ and $A$ are ordinary (or even) differential forms, $G$ and $J$ are usually assumed to be \emph{twisted} (or odd) differential forms~\cite{hehl03, parasecoli05, frankel}. In this context, the volume form is also defined as a twisted form, and the Hodge star operator is hence taken as a map from ordinary forms to twisted forms, and vice versa.
}
\begin{equation}\label{eq:Vacuum}
	G=\hodge F \,.
\end{equation}
The above relation is algebraic in nature, linear, and  strictly local, that is, the value of $G$ at a given spacetime point is related only to the value taken by $F$ at the same point. Let us remark that since the Faraday and Amp\`{e}re equations involve only exterior derivatives, they are covariant under arbitrary diffeomorphisms. On the other hand, the vacuum constitutive equation makes use of the metric through the Hodge star operator. As a result, the symmetries of electrodynamics are the symmetries of the constitutive equation, which are in turn related to the symmetries of the metric.

The distinction between ``electric'' and ``magnetic'', as well as between ``Coulombian'' and ``Amp\`{e}rian'', is brought about by a splitting of spacetime into time and space~\cite{hehl03, preziosi06, CDML19, Fec97}. In the framework developed in the previous section, the time evolution vector field $X_0=\partial/\partial t=T$, generating the one-parameter group of the time translations, allows us to introduce the electric 1-form $E$, and the magnetic 2-form $B$,
\begin{gather}\label{eq:Edef}
	E = i_{T}  F
	=E_x \dd{x} +E_y\dd {y} +E_z \dd {z},
	\\\label{eq:Bdef}
	B = -(  F-\dd{t}\wedge i_{T}F)
	=  B_x\dd {y}\wedge \dd {z}
	+B_y\dd {z}\wedge \dd {x}
	+B_z\dd {x}\wedge \dd {y},
\end{gather}
so that the Faraday form is decomposed as
\begin{equation}\label{eq:Fdef}
	F=\dd{t}\wedge E  - B\,.
\end{equation}
Likewise, the potential 1-form reads $A=\varPhi \dd{t} -A_x\dd{x} -A_y\dd {y} -A_z \dd {z}$.

By the same token, we introduce the magnetic 1-form $H$, and the electric 2-form $D$,
\begin{gather}
	H = -i_{T} G
	= H_x \dd{x} +H_y\dd {y} +H_z \dd {z},
	\\
	D=-(  G-\dd{t} \wedge i_{T} G)
	=D_x\dd {y}\wedge \dd {z}
	+D_y\dd {z}\wedge \dd {x}
	+D_z\dd {x}\wedge \dd {y}\,,
\end{gather}
which decompose the Amp\`{e}re form as
\begin{equation}\label{eq:Gdef}
	G=-\dd{t}\wedge H - D\,.
\end{equation}
Finally, by introducing the current density 2-form $j$, and the charge density 3-form $R$,
\begin{gather}\label{eq:jdef}
	j = i_{T}J
	=j_x\dd{y}\wedge\dd{z}
	+j_y\dd{z}\wedge\dd{x}
	+j_z\dd{x}\wedge\dd{y},
	\\
	R = -(J-\dd{t}\wedge i_{T}J)
	=\rho \dd{x}\wedge\dd{y}\wedge\dd{z},
\end{gather}
the charge-current 3-form reads
\begin{equation}\label{eq:Jdef}
	J=\dd{t}\wedge j - R\,.
\end{equation}
In terms of the above decompositions, the vacuum constitutive equation is equivalent to
\begin{align}\label{eq:EDBH}
	\hodge\qty( \dd{t}\wedge E)= -D\,, \qquad
	\hodge B=\dd{t}\wedge H\,.
\end{align}

Therefore, when Eqs.~\eqref{eq:dF}--\eqref{eq:Vacuum} are read by components, we recover the familiar picture of Eqs.~\eqref{eq:EBhomo}--\eqref{eq:continuity}.

\section{Intrinsic reformulation of the descent}\label{sec:reformulation}
Now we are ready to consider the problem of obtaining a coordinate-free analogue of the procedure of Sec.~\ref{sec:CartesianReduction}. In particular, we would like to recover the separation of electromagnetism into independent equations that we observed in Cartesian coordinates, and we would like such equations to be interpretable as emerging from a low-dimensional spacetime.

In this section we consider the descent to $(2+1)$ of Sec.~\ref{sec:firstdesc}. As noted earlier, descending along one single spatial direction brings about a separation that is analogous to the one occurring in the stationary case. Therefore, it is convenient to first return to this more familiar case, and look at its description in terms of differential forms.
When we consider the electromagnetic theory in a $(3+1)$-dimensional Minkowski spacetime, by requiring fields and sources to be stationary (i.e., the forms $F$, $G$, and $J$ to be invariant under the one-parameter subgroup of time translations), we obtain the mutually independent equations of electrostatics and magnetostatics. 

	The equations of electrostatics are expressed only in terms of $E$ and $D$, those of magnetostatics only in terms of $H$ and $B$. In particular, each set of equations can be  formulated in terms of a 1-form and a 2-form in a 3-dimensional Euclidean manifold. 
	For instance,  starting with the 2-form $\dd{t}\wedge E$ in $(3+1)$ we end up with the 1-form $E$, depending only on the space variables, and $\dd t$ does not appear anymore.

By the same token, the idea that will be explored in the following  is to go from electromagnetism in $(3+1)$ to electromagnetism in $(2+1)$ by imposing invariance of the forms $F$, $G$ and $J$ under the action of a one-parameter subgroup of spatial translations. By the analogy with the stationary case, we expect such condition to provide us with two sets of unrelated $(2+1)$-dimensional equations, each written in terms of  both a 1-form and a 2-form. 

The lesson we learn from the discussion of the stationary case is that 
the implementation of such procedure should allow the electromagnetic forms to ``change type''. In other words, when the dimension of spacetime is lowered, we should allow the differential forms describing our physical quantities to reduce in degree, reflecting the loss of a spatial dimension. For example, a 2-form in four dimensions may yield a 1-form in three dimensions under such a descent.
Let us observe that this feature has no room to emerge for differential equations on scalar functions, such as the scalar wave equation, for the very simple reason that the degree of a $0$-form cannot be lowered. However, when the picture is enlarged to  forms of nonzero degree, such as Maxwell equations, this feature emerges quite naturally, and should be regarded as inherent to the reduction procedure by descent. 

We will develop our discussion by making use of the orthonormal coordinates $(x^\mu)$ introduced in Sec.~\ref{sec:exterior}.
After considering, in Sec.~\ref {sec:scalar}, the descent for scalar functions, we move, in Sec.~\ref {sec:diff}, to the case of differential forms.

\subsection{Scalar functions}\label{sec:scalar}
Let us briefly return on Hadamard's application of the method of descent to the scalar wave equation, summarized in Sec.~\ref{sec:HDMwaveeq}, looking at it from a  geometric perspective. Hadamard starts by observing that the solutions of the homogeneous cylindrical equation are precisely those of the spherical wave equation which do not depend on $z$. Then a cylindrical problem is reduced to the spherical problem with same source term and, possibly, initial data, which are obviously $z$-independent.

First of all, the condition that a smooth scalar function $f\in\mathcal F$ does not depend on $z$, that is, $f=f(t,x,y)$, translates into the (infinitesimal) condition $\partial f/\partial z=0$, that is,
\begin{equation}\label{eq:descScalar}
	\lie f=0\,,
\end{equation}
where $Z = \partial /\partial z$, expressing the fact that $f$ is invariant under translations parallel to the $z$ axis. Such translations form a one-parameter subgroup of the translation group, generated by $Z$, hence are a symmetry of the d'Alembert operator~\eqref{eq:dalembert},
\begin{equation}
	\lie \dal = \dal \lie\,.
\end{equation}

Hadamard's stepping stone is the observation that, if $f$ satisfies the descent condition $\lie f=0$, then the d'Alembert operator~\eqref{eq:dalembert} acts on $f$ as the differential operator $\lie[X_0]\lie[X_0] -\lie[X_1]\lie[X_1] -\lie[X_2]\lie[X_2]$, which may be thought of as a $(2+1)$-dimensional d'Alembert operator. In other words, $\lie f=0$ is sufficient to regard $f$ as arising form a  $(2+1)$-dimensional spacetime (e.g., as a pullback). 
Then, when we study problems for the $(3+1)$-dimensional wave equation where the source term and, possibly, the initial data are invariant under the action of $Z$ (a tangent vector field to the hypersurface $t=0$ of the boundary conditions), their solutions are not searched within the entire algebra $\mathcal{F}$ of smooth functions on spacetime, but within 
\begin{equation}\label{eq:FZ}
	\mathcal F_Z =  
	\qty{\,  f\in \mathcal F
		\; | \; \lie f=0\,}\,,
\end{equation}
the subalgebra of all functions that are invariant under the action of $Z$.

The above procedure is already applicable to electromagnetism in the potential formulation. Choosing the divergence of the potential 1-form $A$ to be uniform (e.g., assuming Lorenz gauge conditions, $\codd A=0$), electromagnetism is described by the wave equation $\dal A=\hodge J$. Then the dynamics of $A$ can be discussed componentwise, and Hadamard's framework can be applied to the scalar wave equations obeyed by each component of the potential 1-form. 

This approach allows us to trace back the procedure followed in Ref.~\cite{descent} to Hadamard's. However, the components of the potential 1-form are treated separately, as if they were generic scalar functions, without paying attention to the geometric object they describe altogether.

In this work we would like to follow a different path, namely, to embrace an intrinsic perspective, and apply Hadamard's \emph{ideology}, rather than his techniques, to the equations of electromagnetism in the field formulation. To do so, we need to understand how to extend the descent procedure to differential forms.

\subsection{Differential forms}\label{sec:diff}
The reduction of electromagnetism to $(2+1)$, described in Sec.~\ref{sec:firstdesc} within the framework of vector calculus, is carried out by assuming that all fields and sources do not depend on $z$, as prescribed by the descent conditions~\eqref{eq:FirstDescCond}. When these quantities are interpreted as differential forms,  
such conditions translate into requiring that all components of $F$, $G$, and $J$ do not depend on $z$, that is, $\lie F=0$, $\lie G=0$, and $\lie J=0$. The applicability of these conditions to the equations of electromagnetism will be the starting point of the next section. 

Here we will discuss the condition
\begin{align}\label{eq:DescOmegaZ}
	\lie \omega =0\,,
\end{align}
on a generic differential form $\omega$, focusing on its implications with respect to dimensional reduction. Particular attention will be paid to $F$, $G$, and $J$. We know that each of these forms brings about \emph{two} forms ---$E$ and $B$, $D$ and $H$, and $R$ and $j$, respectively--- which, in the stationary case, can be interpreted as pullbacks of forms on a $3$-dimensional space. We expect something along these lines when invariance under the action of $Z$, rather than of $T$, is considered.

To begin with, condition~\eqref{eq:descScalar} on scalar functions is just an instance of condition~\eqref{eq:DescOmegaZ} on differential forms. However, while the condition $\lie f=0$ is sufficient to interpret a function $f$ as a pullback of a function on a $(2+1)$-dimensional spacetime, this is not true, in general, for a differential form. 

A $p$-form $\omega$ on $M^4$ can be uniquely written in terms of the global coframe $(\dd{x^\mu})$ as
\begin{align}\label{eq:pform}
	\omega
	=\frac{1}{p!}\,\omega_{\mu_1\dots\mu_p}\,\dd{x^{\mu_1\dots\mu_p}},
\end{align}
where $\dd{x^{\mu_1\dots\mu_p}}= \dd{x^{\mu_1}} \wedge \cdots \wedge\dd{x^{\mu_p}}$ are the basis forms, and all coefficients are assumed to be antisymmetric under exchange of any two indices. Then $\lie \omega =0$ is equivalent to the property that each coefficient function does not depend on $z$. 

However, unless $p=0$, some of the basis form involve the differential $\dd{z}$, which has no counterpart in $(2+1)$ dimensions and thus cannot be represented in a reduced theory unless interpreted as a vector-valued form.
In other words, the condition $\lie\omega=0$ is not enough to interpret a differential form $\omega$ as a pullback of a form on a $(2+1)$-dimensional spacetime: we also need $i_Z\omega=0$. In this respect, the case of scalar functions is simpler, precisely because the latter condition is always satisfied.

Our next task is to single out the quantities that, upon assuming $Z$-invariance, can be regarded as emerging from $(2+1)$. To this end, we only have to look at Eqs.~\eqref{eq:Edef}--\eqref{eq:Jdef}. 

In the stationary case, the forms that can be interpreted as pullbacks from a 3-dimensional space are determined by decomposing $F$, $G$, and $J$ as $\omega= \dd{t}\wedge i_T \omega + (\omega-\dd{t}\wedge i_T \omega)$. This is precisely the decomposition of the exterior algebra induced by $T$ (cf.\ Appendix~\ref{app:projections}). 

The vector field $Z$ induces a completely analogous decomposition: a differential form $\omega$ can be uniquely split as
\begin{equation}\label{eq:decoZ}
	\omega=\dd{z}\wedge \omega_{(1)} + \omega_{(0)}\,,
\end{equation}
upon requiring that $i_Z\omega_{(0)} =0$ and $i_Z\omega_{(1)} =0$; in particular,
\begin{align}\label{eq:decoZ2}
	\omega_{(1)} =i_Z \omega\,, \qquad
	\omega_{(0)} 
	=\omega - \dd{z}\wedge \omega_{(1)} 
	=i_Z(\dd{z}\wedge \omega)\,.
\end{align}
In other words, the exterior algebra $\Omega$ is decomposed into the subalgebra $\Omega_{(0)}=\ker i_Z$ (the $\mathcal F$-algebra generated by $\dd{t}$, $\dd{x}$, and $\dd{y}$), and the submodule $\Omega_{(1)}$
of all forms of the kind $\omega=\dd{z}\wedge \nu$, with $\nu\in \Omega_{(0)}$ (the $\mathcal F$-module of all forms with a factor $\dd{z}$).
If $\omega$ is a $p$-form, $\omega_{(0)}$ is a $p$-form, and $\omega_{(1)}$ is a $(p-1)$-form. 

The decomposition is trivial only on 0-forms, for which $\omega_{(1)} =0$, and on volume forms, for which $\omega_{(0)} =0$. 
Incidentally, observe that $\dd{z}\wedge \omega_{(1)} $ is the sum of all terms in the expansion~\eqref{eq:pform} with a factor $\dd{z}$, and $\omega_{(0)}$ of those without.

Crucially, the condition $\lie\omega=0$ is sufficient to interpret $\omega_{(0)}$ and $\omega_{(1)}$ as pullbacks of forms on a $(2+1)$-dimensional spacetime.
In particular, starting from a $p$-form $\omega$ with $1\le p\le 3$ (as in the case of $F$, $G$, and $J$), the condition $\lie\omega=0$ yields \emph{two} forms emerging from $(2+1)$: the $p$-form $\omega_{(0)}$, and the $(p-1)$-form $\omega_{(1)}$.

Let us give a closer look at the structure of 
\begin{equation}\label{eq:OmegaZ}
	\Omega_Z =  
	\qty{\,\omega\in \Omega 
		\;\middle\vert \; \lie \omega=0\,}\,,
\end{equation}
the subalgebra of all forms that are invariant under the action of $Z$. 
As a result of the decomposition of the exterior algebra, $\Omega_Z$ is decomposed into
\begin{gather}\label{eq:OmegaZ0}
	\Omega_{Z,(0)}
	=\Omega_Z \cap \Omega_{(0)}
	=\qty{\,\omega
		\;\middle\vert \; \lie \omega=0,\; i_Z\omega=0\,}
	\,,
	\\\label{eq:OmegaZ1}
	\Omega_{Z,(1)}
	=\Omega_Z \cap \Omega_{(1)}
	= \qty{\,\dd{z}\wedge \nu
		\;\middle\vert \; \lie \nu=0,\; i_Z\nu=0\,}
	\,.
\end{gather}
In other words $\Omega_{Z,(0)}$ (the $\mathcal{F}_Z$-algebra generated by $\dd{t}$, $\dd{x}$, and $\dd{y}$) is the subalgebra of all forms that can be interpreted as pullbacks from $(2+1)$, whereas $\Omega_{Z,(1)}$ (the $\mathcal F_Z$-module of all forms with a factor $\dd{z}$) is the submodule of all forms of the kind $\dd{z}\wedge \nu$, where $\nu = i_Z\omega$ can be interpreted as a pullback from $(2+1)$.

This observation suggests the idea of interpreting the forms in $\Omega_{Z,(0)}$ as ordinary forms, and the forms in $\Omega_{Z,(1)}$ as vector-valued forms, ``with value $\dd{z}$''. For instance, we can use an ``abstract'' vector $\vb{e}_z$ as a ``placeholder'' for the 1-form $\dd{z}$, consider the exterior algebra $\Lambda (V)$ of the one-dimensional real vector space $V=\Span \qty{\vb{e}_z}$, and define the map
\begin{align}\label{eq:vectorval}
	\Omega_{Z,(0)}\ni\omega\mapsto  \omega\otimes 1\,, 
	\qquad
	\Omega_{Z,(1)}\ni\omega\mapsto i_Z\omega\otimes \vb{e}_z\,,
\end{align}
of $\Omega_Z=\Omega_{Z,(0)}\oplus \Omega_{Z,(1)}$ to the space of the $\Lambda (V)$-valued differential forms. In this way, $\Omega_{Z,(0)}$ and $\Omega_{Z,(1)}$ are mapped to $\Lambda^0 (V)$- and $\Lambda^1(V)$-valued forms, respectively.

In terms of coordinates,  a $Z$-invariant $p$-form $\omega$ reads
\begin{align}\label{eq:pformZ}
	\omega
	= \frac{1}{p!}\,
	\omega_{\mu_1\dots\mu_p}(t,x,y)\,\dd{x^{\mu_1\dots\mu_p}},
\end{align}
with $\mu_i=0,1,2,3$,
and its components with respect to the decomposition~\eqref{eq:decoZ} are
\begin{gather}\label{eq:vector}
	\dd{z}\wedge\omega_{(1)}
	=\dd{z}\wedge i_Z\omega
	=\dd{z}\wedge\frac{1}{(p-1)!}
	\sum_{\mu_k\ne 3}
	\omega_{3\mu_1\dots\mu_{p-1}}(t,x,y)\,\dd{x^{\mu_1\dots\mu_{p-1}}},
	\\\label{eq:scalar}
	\omega_{(0)}
	=\omega-\dd{z}\wedge\omega_{(1)}
	=\frac{1}{p!}\,
	\sum_{\mu_k\ne 3}
	\omega_{\mu_1\dots\mu_p}(t,x,y)\,\dd{x^{\mu_1\dots\mu_p}}.
\end{gather}
The dependence on the coordinates has been made explicit to underline that all coefficient functions in the above expressions are in $\mathcal{F}_Z$.
Then $\omega$ is decomposed into a \emph{scalar component}, $\omega_{(0)}\cong\omega_{(0)} \otimes 1$, which can be interpreted as a 
a pullback of an ordinary $p$-form in $(2+1)$, and a \emph{vector component}, $\dd z\wedge\omega_{(1)}\cong \omega_{(1)}\otimes \vb e_z$, which can be interpreted as a pullback of a vector-valued $(p-1)$-form in $(2+1)$.

\subsection{Application to electromagnetism}\label{eq:descent}
In this section, the framework developed so far will be applied to electromagnetism. As a first task, we need to make sure that the descent conditions are implemented correctly.
Essentially, since the vacuum constitutive equation~\eqref{eq:Vacuum} is a constraint on $F$ and $G$, we are not allowed to \emph{independently} assume $\lie F=0$, and $\lie G=0$. 

In other words, we are led to consider how the Lie derivative of a form, and of its Hodge dual, are related to each other. In our settings, this problem is rather simple: it is sufficient to observe that any nontrivial basis form $\dd{x^{\mu_1\dots\mu_p}}$ has a vanishing Lie derivative with respect to $Z$, and that its Hodge dual is equal, up to a sign, to another nontrivial basis form. As a result, the identity
\begin{equation}\label{eq:commu}
	\lie \hodge\omega=\hodge \lie \omega\,
\end{equation}
holds on all $p$-forms of the kind $\omega=f \dd{x^{\mu_1\dots\mu_p}}$ (by Leibniz rule), and then extends to all forms (by linearity).\footnote{
	In the general case, the commutator of a Lie derivative and the Hodge star operator is nontrivial~\cite{trautman}.
	} 
As a consequence, $\Omega_Z$ in~\eqref{eq:OmegaZ} is invariant under the Hodge star operator, and, in particular, the conditions $\lie F=0$ and $\lie G=0$ are equivalent.

In accordance with our reinterpretation of Hadamard's procedure, the reduction of electromagnetism by descent will be performed by assuming
\begin{align}\label{eq:Desc}
	\lie F=0\,, \qquad \lie G=0\,, \qquad \lie J=0\,,
\end{align}
that is, by restricting $F$, $G$, and $J$ to $\Omega_Z$.

\subsubsection{Reduction of the equations}
Let us find out the consequences of conditions~\eqref{eq:Desc} by considering each equation separately. 
To this end, observe by Eq.~\eqref{eq:decoZ} that
\begin{equation}\label{eq:decodZ}
	\dd{\omega}=-\dd{z}\wedge\dd{\omega_{(1)}} + \dd{\omega_{(0)}}\,.
\end{equation}
Furthermore, we always have $\lie\dd{\omega_{(r)}} =\dd{\lie\omega_{(r)}}=0$, and, upon assuming $\lie\omega=0$, we also have $i_Z \dd{\omega_{(r)}}=-\dd{i_Z\omega_{(r)}}=0$. As consequence, when $\omega$ is $Z$-invariant, the decomposition~\eqref{eq:decoZ} of $\dd{\omega}$ is just given by~\eqref{eq:decodZ}.

The condition $\lie F=0$ splits the Faraday equation~\eqref{eq:dF}, $\dd{F}=0$, into two equations, in terms of the scalar and the vector components of $F$,
\begin{gather}
	\label{eq:FaraEEB}
	\dd  F_{(0)}=0\,,\\
	\label{eq:FaraBBE}
	\dd  F_{(1)}=0\,.
\end{gather}
Since $F_{(0)}$ and $F_{(1)}$ can be thought of as pullbacks of $(2+1)$-dimensional forms, the above equations can be interpreted ``before the pullback'', that is, as two separate equations in $(2+1)$, in terms of two different kinds of objects (a 2- and a 1-form, respectively).

By the same token, the condition $\lie G=0$ splits the Amp\`{e}re equation~\eqref{eq:dG}, $\dd{G}=J$, into two equations, in terms of the scalar and the vector components of $G$ and $J$,
\begin{gather}
	\label{eq:ampereBBE}
	\dd G_{(0)}=J_{(0)}\,, 
	\\
	\label{eq:ampereEEB}
	\dd  G_{(1)}= -J_{(1)}\,.
\end{gather}
As before, the above equations can be given a novel interpretation in $(2 + 1)$, and relate objects of different kinds: Eq.~\eqref{eq:ampereBBE} involves a 2- and a 3-form, Eq.~\eqref{eq:ampereEEB} a 1- and a 2-form.

As for the continuity equation~\eqref{eq:dJ}, $\lie J=0$ splits $\dd{J}=0$ into $\dd J_{(0)}=0$, and $\dd J_{(1)}=0$. However, in all generality (i.e., without assuming $\lie J=0$), we have
\begin{equation}
	\dd{J_{(0)}}=\dd{i_Z(\dd{z}\wedge J)}
	=\lie (\dd{z}\wedge J) -i_Z \dd(\dd{z}\wedge J)
	=\dd{z}\wedge \lie J 
	=\dd{z}\wedge \lie J_{(0)}\,,
\end{equation}
so that $\dd J_{(0)}=0$ is equivalent to $\lie J_{(0)}=0$. As a consequence, the continuity equation yields just one nontrivial reduced equation,
\begin{equation}
	\dd  J_{(1)}=0\,,
\end{equation}
which, once again, can be thought in $(2+1)$.

Let us finally turn to the constitutive equation. 
Here we need a metric property of the decomposition~\eqref{eq:decoZ}, which is proved in Appendix~\ref{app:compatibility}: the duality relation $\nu=\hodge \omega$ is equivalent to the identities
\begin{align}\label{eq:hodgedecomp}
	\dd{z}\wedge  \nu_{(1)}=\hodge  \omega_{(0)}\,, 
	\qquad 
	\nu_{(0)}=
	\hodge \qty(\dd{z}\wedge\omega_{(1)})\,.
\end{align}
If $\omega$ is a $p$-form, the above relations can also be put in the form of identities between objects of $\Omega_{(0)}=\ker i_Z$ (which do not involve the 1-form $\dd{z}$), as
\begin{align}\label{eq:decoHodge}
	\nu_{(1)}=i_Z\hodge\omega_{(0)}\,, \qquad 
	\nu_{(0)}=(-1)^p\,i_Z\hodge\omega_{(1)}\,.
\end{align}
In other words, the Hodge operator $\hodge$ is off-diagonal with respect to the decomposition~\eqref{eq:decoZ} induced by $Z$, as $\Omega_{(0)}$ and $\Omega_{(1)}$ are mapped into each other (hence, when we consider $Z$-invariant forms, scalar forms are mapped to vector forms, and vice versa).

In particular, the constitutive equation $G=\hodge F$ is equivalent to the identities
\begin{align}\label{eq:HodgeGsv}
	G_{(1)}=i_Z\hodge F_{(0)}\,,
	\qquad 
	G_{(0)}=i_Z\hodge F_{(1)}\,,
\end{align}
which are analogous to the relations $H=i_T \hodge B$, and $D=i_T\hodge E$, rephrasing Eq.~\eqref{eq:EDBH}. Such equivalence holds in full generality. However, once we assume that $ F$ and $G$ are $Z$-invariant, the analogy with the stationary case suggests to interpret the equations above as representing two $(2+1)$-dimensional duality relations.

\subsubsection{Descent to  \texorpdfstring{$(2+1)$}{(2+1)}}
The constitutive equation directs us on how to put the reduced equations together: from Eq.~\eqref{eq:HodgeGsv} we learn, in the first place, that Eq.~\eqref{eq:FaraEEB} should be paired with Eq.~\eqref{eq:ampereEEB}, and Eq.~\eqref{eq:FaraBBE} with Eq.~\eqref{eq:ampereBBE}. As a result, we end up with two sets of reduced equations: one involving $F_{(0)}$, $G_{(1)}$, and $ J_{(1)}$,
\begin{gather}\label{eq:newEEB1}
	\dd F_{(0)}=0\,,
	\\
	\label{eq:newEEB2}
	\dd   G_{(1)}= - J_{(1)}\,, 
	\\
	\label{eq:newEEB3}
	G_{(1)}= i_Z \hodge F_{(0)}\, ,
\end{gather}
and the other  involving $F_{(1)}$, $G_{(0)}$, and $J_{(0)}$,
\begin{gather}\label{eq:newBBE1}
	\dd  F_{(1)}=0\,,
	\\\label{eq:newBBE2}
	\dd G_{(0)}=J_{(0)}\,,
	\\\label{eq:newBBE3}
	G_{(0)}= i_Z\hodge  F_{(1)}\,.
\end{gather}

Since $F_{(r)}$, $G_{(r)}$, and $J_{(r)}$ can be thought of as pullbacks of forms on a $(2+1)$-dimensional spacetime, each of the above set can be conceived in $(2+1)$. Then, it's easy to see that Eqs.~\eqref{eq:newEEB1}--\eqref{eq:newEEB3} are a reformulation of the EEB equations~\eqref{eq:EEB1}--\eqref{eq:EEB4}, and  Eqs.~\eqref{eq:newBBE1}--\eqref{eq:newBBE3} of the BBE equations~\eqref{eq:BBE1}--\eqref{eq:BBE4}. All results of Sec.~\ref{sec:firstdesc} have been recovered.

Let us conclude with some observations.
First of all, within the low-dimensional framework, there is no trace of the fact that $F_{(0)}$ and $F_{(1)}$ both come from $F$, $G_{(0)}$ and $G_{(1)}$ from $G$, and $J_{(0)}$ and $J_{(1)}$ from $J$. To recover the single equation $G=\hodge F$ in $(3+1)$, the $(2+1)$-dimensional relations~\eqref{eq:newEEB3} and~\eqref{eq:newBBE3} must be ``unified'', and such unification follows from reinstalling the 1-form $\dd{z}$ precisely where it has been removed. Once the $(3+1)$-dimensional objects are reconstructed, Eqs.~\eqref{eq:newEEB1} and~\eqref{eq:newBBE1} can be put together into Faraday equation, and Eqs.~\eqref{eq:newEEB2} and~\eqref{eq:newBBE2} into Amp\`{e}re equation, resulting in the ``unified'' theory of electromagnetism in $(3+1)$.

Furthermore, two homogeneous forms on an odd-dimensional manifold that are related by Hodge duality  are necessarily of different degree. Therefore, if we assume that electromagnetism in $(2+1)$ should have the same structure as in $(3+1)$ (i.e., should consist of one homogeneous equation, and one inhomogeneous, in terms of dual quantities), the only suitable way to put the reduced equations together is the one we derived from Eq.~\eqref{eq:HodgeGsv}.

Finally, from the analogy with the stationary case, we expected that $Z$-invariance would separate the equations of electromagnetism into two sets of unrelated $(2+1)$-dimensional equations, each written in terms of  both a 1-form and a 2-form. Indeed, the picture we have obtained precisely mirrors the separation of stationary electromagnetism into electrostatics and magnetostatics. In other words, following Ref.~\cite{descent}, one could interpret stationary electromagnetism as a special case of the descent procedure. The generator of time translations, $T=\partial/\partial t$, induces a decomposition of the exterior algebra allowing us to write $F$, $G$, and $J$ in terms of forms that coincide, up to a sign, with $E$, $B$, $H$, $D$, $R$, and $j$. The assumption of $T$-invariance on the one hand allows us to interpret these forms as $3$-dimensional, and, on the other, determines a separation of the equations along such decomposition. Since the Hodge operator in $(3+1)$ is off-diagonal also with respect to the decomposition induced by $T$, the constitutive equation tell us how to put together the reduced equations. In the end, the stationary equations in $(3+1)$ can be seen as pullbacks of two separate theories, electrostatics and magnetostatics, given in terms of 2 equations, in a 1- and a 2-form, related by 3-dimensional Hodge duality. In the 3-dimensional framework, the relations between electric and magnetic quantities, as well as between static charges and stationary currents, as components of the same $(3+1)$-dimensional form, is lost.

\section{Multiple descent}\label{sec:seconddescent} 

\subsection{Multiple descent conditions}
\label{sec:multiple}

Let us now consider the coordinate vector field $Y=\partial/\partial y$, giving rise to a decomposition of the exterior algebra which is completely analogous to the one determined by $Z$. Since $[Y,Z]=0$ one can implement a double descent sequentially in any order.

An arbitrary differential form $\omega$ is decomposed in four components:
\begin{align}\label{eq:decoMult}
	\omega
	= \omega_{(0,0)}+\dd y \wedge\omega_{(1,0)}+\dd z \wedge\omega_{(0,1)}+\dd y \wedge \dd z \wedge\omega_{(1,1)}\,,
\end{align}
with 
\begin{equation}
	 i_Y \omega_{(r,s)} = i_Z \omega_{(r,s)}=0, \qquad r,s\in\{0,1\}. 
\end{equation}
Here we have
\begin{gather}
	\omega_{(0,0)}=-i_Y i_Z (\dd{y}\wedge\dd{z}\wedge \omega)\,,\\
	\omega_{(0,1)}=-i_Y i_Z (\dd{y}\wedge\omega)\,,\\
	\omega_{(1,0)}=i_Y i_Z (\dd{z}\wedge\omega)\,,\\
	\omega_{(1,1)}=-i_Y i_Z \omega\,.
\end{gather}

Next, we assume that our $p$-form $\omega$ is both $Y$- and $Z$-invariant, that is
\begin{align}\label{eq:DescOmegaYZ}
	\lie \omega=0\,,\qquad\lie[Y] \omega=0\,.
\end{align}
The above \emph{multiple descent} conditions are respectively equivalent to $\lie[Z]\omega_{\mu_1\dots\mu_p}=0$ and $\lie[Y]\omega_{\mu_1\dots\mu_p}=0$, hence the coordinate expression~\eqref{eq:pform} reads
\begin{align}\label{eq:pformYZ}
	\omega
	=\frac{1}{p!}\,
	\omega_{\mu_1\dots\mu_p}(t,x)\,\dd{x^{\mu_1\dots\mu_p}}\,, \qquad \omega_{\mu_1\dots\mu_p}\in\mathcal F_{YZ}\,,
\end{align}
where by
\begin{equation}\label{eq:FYZ}
	\mathcal F_{YZ} 
	= \mathcal F_{Y}\cap \mathcal F_{Z} 
	=\qty{\,  f\in \mathcal F
		\;\middle\vert \; \lie[Y] f=0,\,\lie f=0\,}
\end{equation}
we denote the subalgebra of smooth functions that are both $Y$- and $Z$-invariant. Also in this case the set $\Omega_{YZ}=\Omega_{Y}\cap \Omega_{Z}$ of all the $Y$- and $Z$-invariant forms, the $\mathcal F_{YZ}$-mo\-du\-le generated by  $\dd{t}$, $\dd{x}$, $\dd{y}$ and $\dd{z}$, is a subalgebra of $\Omega$ and is decomposed as $\Omega_{YZ}=\bigoplus_p\Omega^p_{YZ}$.
When the decomposition in Eq.~\eqref{eq:decoMult} is specialized for a $Y$- and $Z$-invariant homogeneous $p$-form $\omega$, we 
obtain the following explicit expressions:
\begin{gather}
	\label{eq:QYQZ}
	\omega_{(0,0)}
	=\frac{1}{p!}\,
	\sum_{\mu_k\ne 2,3}
	\omega_{\mu_1\dots\mu_p}(t,x)
	\,\dd{x^{\mu_1\dots\mu_p}}\,,\\
	\label{eq:PYQZ}
	\omega_{(1,0)}
	=\frac{1}{(p-1)!}
	\sum_{\mu_k\ne 2,3}
	\omega_{2\mu_1\dots\mu_{p-1}}(t,x)
	\,\dd{x^{\mu_1\dots\mu_{p-1}}} \,,\\
	\label{eq:QYPZ}
	\omega_{(0,1)}
	=\frac{1}{(p-1)!}
	\sum_{\mu_k\ne 2,3}
	\omega_{3\mu_1\dots\mu_{p-1}}(t,x)
	\,\dd{x^{\mu_1\dots\mu_{p-1}}}\,,\\
	\label{eq:PYPZ}
	\omega_{(1,1)}=
	\frac{1}{(p-2)!} 
	\sum_{\mu_k\ne 2,3}
	\omega_{23\mu_1\dots\mu_{p-2}}(t,x)
	\,\dd{x^{\mu_1\dots\mu_{p-2}}}
	\,.
\end{gather}
 
While the component $\omega_{(0,0)}$ can be immediately thought of as the pullback of a $p$-form, $\dd y \wedge\omega_{(1,0)}$ and $\dd z \wedge\omega_{(0,1)}$ can be conceived as pullback of $(p-1)$-forms only after a contraction with $Y$ and $Z$, respectively, and $\dd y \wedge\dd z \wedge\omega_{(1,1)}$ only after a contraction with  $Y\wedge Z$. Then, by considering the exterior algebra  $\Lambda(V)$ of the 2-dimensional real vector space $V=\Span \qty{\vb{e}_y,\vb{e}_z}$, each component in Eqs.~\eqref{eq:QYQZ}--\eqref{eq:PYQZ} can be mapped  as before  to a different kind of geometric object. Namely, we have a scalar component
\begin{align}
	\omega_{(0,0)}\mapsto \omega_{(0,0)}\otimes 1\,, 
\end{align}
two vector-valued components of degree 1,
\begin{align}
	\dd y \wedge\omega_{(1,0)}\mapsto \omega_{(1,0)}\otimes \vb{e}_y\,, \qquad
	\dd z \wedge\omega_{(0,1)}\mapsto \omega_{(0,1)}\otimes \vb{e}_z\,,
\end{align}
and one vector-valued component of degree 2 (a \emph{pseudoscalar/bi-vector}),
\begin{align}
	\dd y \wedge\dd z \wedge\omega_{(1,1)}\mapsto  \omega_{(1,1)}\otimes \vb{e}_y\wedge\vb{e}_z\,. 
\end{align} 
In this way, all differential forms which are invariant with respect to both $Y$ and $Z$ can be interpreted as pullbacks of different $\Lambda(V)$-valued differential forms in $(1+1)$.

\subsection{Application to electromagnetism} 
By proceeding on the same lines of Sec.~\ref{eq:descent}, we  find that assuming the descent conditions  $\lie[Y] F=\lie[Z] F=0$, i.e.\ requiring that $F\in\Omega_{YZ}$, ensures that the Faraday equation $\dd F=0$ decomposes in four equations,
\begin{gather}
	\dd F_{(0,0)}=0\,,\label{eq:dF00}\\
	\dd  F_{(1,0)}=0\,,\label{eq:dF10}\\
	\dd  F_{(0,1)}=0\,,\label{eq:dF01}\\
	\dd  F_{(1,1)}=0\label{eq:dF11}\,,
\end{gather}
where the quantities $F_{(0,0)}$, $ F_{(1,0)}$, $ F_{(0,1)}$ and $F_{(1,1)}$ can be conceived as the pullback of different forms in $(2+1)$, respectively of degree $2$, $1$, $1$, and $0$. In particular, by observing that in the orthonormal coframe $\qty(\dd{x^\mu})$ we have that
\begin{gather}
	F_{(0,0)}=E_x\dd{t}\wedge \dd{x}\,,\\
	F_{(1,0)}=-E_y\dd{t} +B_z\dd{x}\,,\\
	F_{(0,1)}=-E_z\dd{t}-B_y\dd{x}\,,\\
	F_{(1,1)}=-B_x\,,
\end{gather}
we easily conclude that Eq.~\eqref{eq:dF00} is trivial (giving $\partial_y E_x=\partial_zE_x=0$), Eq.~\eqref{eq:dF10} coincides with the homogeneous equation of the sector $(E_y,B_z)$, that is $\partial_t B_z+\partial_x E_y=0$, Eq.~\eqref{eq:dF01} coincides with the homogeneous equation of the sector $(B_y,E_z)$, that is $\partial_t B_y-\partial_x E_z=0$, whereas Eq.~\eqref{eq:dF11} gives both the equations of the sector $(B_x)$, that is $\partial_x B_x=0$ and $\partial_t B_x=0$.

By assuming that $G\in \Omega_{YZ}$ and $J\in \Omega_{YZ}$, the Amp\`{e}re equation decomposes as
\begin{gather}
	\dd G_{(0,0)}=J_{(0,0)}\,,\label{eq:dG00}\\
	\dd  G_{(1,0)}=- J_{(1,0)}\,,\label{eq:dG10}\\
	\dd  G_{(0,1)}=-J_{(0,1)}\,,\label{eq:dG01}\\
	\dd  G_{(1,1)}=  J_{(1,1)}\label{eq:dG11}\,.
\end{gather}
At this point, by exploiting Eq.~\eqref{eq:hodgedecomp} 
one gets that the identity $\nu=\hodge\omega$ is equivalent to the following four identities:
\begin{gather}
	\nu_{(1,1)}=i_{Y\wedge Z}\hodge\omega_{(0,0)}\,,\\ 
	\nu_{(0,1)}=(-1)^{|\omega|}i_{Y\wedge Z}\hodge\omega_{(1,0)}\,,\\
	\nu_{(1,0)}=(-1)^{|\omega|-1}i_{Y\wedge Z}\hodge\omega_{(0,1)}\,,\\
	\nu_{(0,0)}=i_{Y\wedge Z}\hodge \omega_{(1,1)}\,,
\end{gather}
where $i_{Y\wedge Z}=i_Z i_Y$ and $|\omega|$ is the degree of $\omega$.
By specializing the above relations to the constitutive equation $G=\star F$, and using Eqs.~\eqref{eq:dG00}--\eqref{eq:dG11}, we finally obtain that the equation $\dd\hodge F=J$ decomposes as
\begin{gather}
	\dd i_{Y\wedge Z}\hodge F_{(1,1)}=   J_{(0,0)}\,,\label{eq:dstarF11}\\
	\dd i_{Y\wedge Z} \hodge F_{(0,1)}=  J_{(1,0)}\,,\label{eq:dstarF01}\\
	\dd i_{Y\wedge Z} \hodge F_{(1,0)}= - J_{(0,1)}\,,\label{eq:dstarF10}\\
	\dd i_{Y\wedge Z} \hodge F_{(0,0)}=  J_{(1,1)} \label{eq:dstarF00}\,.
\end{gather}
Here Eq.~\eqref{eq:dstarF11} is trivial (giving $\partial_y B_x=\partial_z B_x=0$), 
Eq.~\eqref{eq:dstarF01} and Eq.~\eqref{eq:dstarF10} coincide respectively with the non-homogeneous equations of the sectors $(B_y,E_z)$ and $(E_y,B_z)$, namely $-\partial_t E_z + \partial_x B_y = j_z$ and $-\partial_t E_y - \partial_x B_z = j_y$, whereas Eq.~\eqref{eq:dstarF00} gives both the equations of the sector $(E_x)$, that is $\partial_x E_x = \rho$ and $-\partial_t E_x = j_x$.

To sum up, we find that the sector $(E_x)$ is described only by the equation
\begin{equation}
	\dd i_{Y\wedge Z} \hodge F_{(0,0)}=  J_{(1,1)}=0 \,,
\end{equation}
associated with the scalar component $F_{(0,0)}$, the sectors $(E_y,B_z)$ and $(B_y,E_z)$ are described, respectively, by the coupled equations
\begin{align}
	\dd  F_{(1,0)}=0\,,  \qquad   \dd i_{Y\wedge Z} \hodge F_{(1,0)}=  J_{(0,1)}\,,
\end{align}
and 
\begin{align}
	\dd   F_{(0,1)}=0\,, \qquad    \dd i_{Y\wedge Z} \hodge F_{(0,1)}=- J_{(1,0)}\,,
\end{align}
associated with the vector components $F_{(1,0)}$ and $F_{(0,1)}$,
whereas the  sector $(B_x)$ is described only by the equation
\begin{equation}
	\dd F_{(1,1)}=0
\end{equation} 
associated with the pseudoscalar/bi-vector component $F_{(1,1)}$.

\section{Conclusions}

We  developed a coherent geometric framework for the dimensional reduction of classical electromagnetism through the method of descent along vector fields commuting with the wave operator. By exploiting the relationship between the Lie derivative and the Hodge star operator, we established the correct implementation of descent conditions on differential forms that ensures consistency of the constitutive relations under reduction.
Our analysis shows that imposing invariance under a chosen vector field naturally decomposes Maxwell's equations into two decoupled sets of lower-dimensional equations, which describe distinct electromagnetic theories. This reduction procedure yields not only the standard theory, but also an alternative electromagnetic theory involving vector-valued differential forms.
Furthermore, the framework extends naturally to multiple descent along commuting vector fields, resulting in a finer decomposition of Maxwell's equations into sectors involving scalar, vector and bi-vector forms. This richer structure may prove particularly useful for exploring electromagnetic phenomena in lower-dimensional spacetimes and in systems with multiple symmetries. It would be also interesting to explore the effects of dimensional reduction on different constitutive relations~\cite{hehl03,Punti97}, and on topological field theories including a Chern--Simons form.
Although we did not discuss the explicit construction of the reduced carrier space, since it lies beyond the scope of our analysis, we note that the latter is usually identified with the orbit space of a given group action~\cite{raumonen11,nikolov12}, which, depending on the specific case considered, may or may not possess a smooth differential structure. 
Overall, our results provide a unified, geometrically transparent interpretation of dimensional reduction of electromagnetism, clarifying its connection to classical electrostatics and magnetostatics, and paving the way for further investigations of gauge theories and Dirac-type equations in pseudo-Riemannian manifolds and parallelizable Bianchi universes.

\section*{Acknowledgements}
We acknowledge support from INFN through the project ``QUANTUM'' and from the Italian National Group of Mathematical Physics (GNFM-INdAM). 
EE, PF and FVP acknowledge support from PNRR MUR project CN00000013-``Italian National Centre on HPC, Big Data and Quantum Computing''. 
RM and SP acknowledge  support from the PNRR MUR project PE0000023-NQSTI. 
GA acknowledges support from MUR-Italian Ministry of University and Research and Next Generation EU within PRIN 2022AKRC5P ``Interacting Quantum Systems: Topological Phenomena and Effective Theories''.
PF, RM, SP and FVP acknowledge support from the Italian funding within the ``Budget MUR - Dipartimenti di Eccellenza 2023--2027''  - Quantum Sensing and Modelling for One-Health (QuaSiModO).	

\appendix

\section{Laplace--Beltrami and differential operators} 
This appendix is structured as follows. In Sec.~\ref{sec:prelim} we review some basic notions on pseudo-Riemannian manifolds. In particular, we recall how the Hodge star operator, the codifferential, and the Laplace--Beltrami operator can be introduced on an oriented pseudo-Riemannian manifold. In Sec.~\ref{sec:differential} we review some basic notions on differential operators, observing how a second order operator with a nondegenerate principal symbol naturally defines a metric. As a result, if the manifold is also orientable, such operator induces a Laplace--Beltrami operator, and we can compare the two actions on differential forms. In Sec.~\ref{sec:DaleVsLB} we then give sufficient conditions under which two  homogeneous differential operators act in the same way.

\subsection{Pseudo-Riemannian manifolds}\label{sec:prelim}
A pseudo-Riemannian metric $g$ on an $m$-dimensional manifold $\mathcal M$ is a symmetric and non-degenerate $(0,2)$-tensor field on $\mathcal M$. 
A metric induces a vector bundle isomorphism between the tangent and the cotangent bundles, determined point-wise by 
\begin{align}\label{eq:flat} 
	T_p\mathcal M\ni v\mapsto v^\flat=g_p(v,\cdot) \in T_p^\ast\mathcal M\,,
	&&
	T_p^\ast\mathcal M\ni \varphi\mapsto \varphi^\sharp=\tilde g_p(\varphi,\cdot) \in T_p\mathcal M\,,
\end{align}
where the $(2,0)$-tensor field $\tilde g$, in turn symmetric and non-degenerate, is the \emph{contravariant form} of the (covariant) metric $g$.
On the one hand, these isomorphisms extends to arbitrary tensor bundles. In particular, we have a $\mathcal F(\mathcal M)$-module isomorphism between vector fields and 1-forms,
\begin{align}\label{eq:VectorfieldsVsForms} 
	\mathfrak{X}\qty(\mathcal M)\ni X\mapsto X^\flat \in\Omega^1\qty(\mathcal M) \,,
	&&
	\Omega^1\qty(\mathcal M)\ni \alpha\mapsto \alpha^\sharp \in\mathfrak{X}\qty(\mathcal M)
	\,,
\end{align}
as well as
\begin{equation}
	\tilde g\qty(\alpha^1\wedge\dots\wedge\alpha^k,
	\beta^1\wedge\dots\wedge\beta^k) 
	=\det 
	(\tilde g(\alpha^i,\beta^j))_{1\le i,j\le k}\,,
\end{equation}
for $\alpha^j,\beta^j$ 1-forms, and $(\omega\wedge\nu)^\sharp = \omega^\sharp\wedge \nu^\sharp$~\cite{flanders,thirring,hehl03}.

In general a manifold is orientable if and only if there exist a nowhere vanishing top-degree form. Moreover, if the manifold is orientable, any nowhere vanishing top-degree form and its opposite determine opposite orientations (hence nowhere vanishing top-degree forms are also called orientation forms). Now, let our pseudo-Riemannian manifold $\mathcal M$ be orientable. Then $\tilde g(\omega,\omega)$ has the same sign for all orientation forms $\omega$, equal to $(-1)^s$, the parity of the negative index of inertia $s$ of $g$, a characteristic of the metric. In particular, there exist precisely two orientation forms, opposite of each other, such that $\tilde g(\omega,\omega)=(-1)^s$.

Now let $\mathcal M$ be oriented. The unique positively oriented normalized orientation form will be called \emph{metric volume form}, and denoted by $\Omega_g$.
The \emph{Hodge star operator} is then defined by
\begin{equation}\label{eq:hodge}
	\hodge \omega  =  i_{\omega^\sharp} \Omega_g\,,
\end{equation}
where $i_{Y_1\wedge\dots\wedge Y_k}=i_{Y_k} \cdots  i_{Y_1}$, from which follows the identity
\begin{equation}\label{eq:InuHodge}
	\hodge (\omega\wedge \nu)=i_{\nu^\sharp}\hodge \omega\,.
\end{equation}
Specifically, $p$-forms are isomorphically mapped to $(m-p)$-forms, as the following \emph{duality property} holds:
\begin{align}\label{eq:HodgeDuality}
	\hodge \hodge \omega  =  (-1)^{p(m-p)+s}\omega \,,
	\qquad
	\omega \in \Omega^p(\mathcal M)\,.
\end{align}
Observe that the metric is playing a two-fold role in the definition of the Hodge star operator, entering both the volume form, through its normalization, and the contraction, through the isomorphism between covariant and contravariant antisymmetric tensors.
The \emph{codifferential} is defined by its action on homogeneous forms,
\begin{align}\label{eq:codiff}
	\codd \omega =(-1)^{m(p+1)+s+1} \,\hodge \dd{} \hodge \omega\,, \qquad \omega \in \Omega^p(\mathcal M)\,,
\end{align}
where we are following the sign convention of~\cite{beppe}.\footnote{
		Let us account for this sign convention. First, $\mathcal M$ can be equipped with a square-integrable inner product, defined on $p$-forms by $\left( \nu, \omega\right) = \int_{\mathcal M}  \nu\wedge\hodge \omega = \int_{\mathcal M}  g( \nu, \omega) \Omega_g$, the measure being determined by the metric volume form. Then the codifferential is introduced as adjoint to the exterior derivative, namely, $\left( \nu,\codd\omega\right)=\left( \dd{\nu},\omega\right)$, for $\omega$ and $\nu$ homogeneous of degree $p$ and $p-1$, respectively --- in these considerations, $\mathcal M$ is \emph{not} a manifold with boundary. Thus the sign in Eq.~\eqref{eq:codiff} ultimately follows from to the duality property~\eqref{eq:HodgeDuality}. Some authors adopt the opposite convention~\cite{thirring}.}
It is a linear map on the exterior algebra, vanishing on functions, and homogeneous of degree $-1$, that is mapping $p$-forms to $(p-1)$-forms. It squares to zero,
but is not a derivation, as Leibniz rule does not hold (for instance, if $f$ is a  function, and $\omega$ is a form of degree $p>0$, in general $\codd (f\omega )\ne f \,\codd\omega$. 
The differential and the codifferential together allow us to construct the \emph{Laplace--Beltrami operator}, $\dd\codd+\codd\dd=(\dd+\codd)^2$, a linear map on the exterior algebra which is homogeneous of order 0. The sign convention we are adopting for the codifferential ensures that the Laplace--Beltrami operator on $\mathbb R^n$ with the Euclidean metric acts on functions the opposite of the Laplacian~\cite{beppe}.

\subsection{Differential operators}\label{sec:differential}
We recall here some notions on linear differential operators. If $f$ is a smooth function on a manifold $\mathcal M$, let $\hat f$ denote the multiplication operator by $f$, that is, the linear map $g\mapsto f g$ on the algebra $\mathcal{F}(\mathcal M)$ of smooth functions. A \emph{linear differential operator} of order at most $k$ on $\mathcal M$ is a linear map $D$ on $\mathcal{F}(\mathcal M)$ such that
\begin{align}
	\comm{\comm{\dots \comm{D}{\hat f_{k+1}}}{\dots\,,\hat f_2}}{\hat f_1}=0\,,
\end{align}
for all $f_1,\dots, f_{k+1}\in \mathcal {F}(\mathcal M)$, where $\left[\,\cdot\,,\cdot\,\right]$ is the commutator of linear maps on $\mathcal{F}(\mathcal M)$. Then a linear differential operator is of order $k$ if its order is at most $k$ but not at most $k-1$. Differential operators of order $0$ are all and only the multiplication operators by a smooth function, whereas homogeneous differential operators of order 1 are all and only the vector fields.
The \emph{principal symbol} of a differential operator $D$ of order $k$ on $\mathcal M$ is the contravariant $k$-tensor field on $\mathcal M$ determined by
\begin{equation}
	\qty(\sigma(D) )\qty(\dd f_1,\dots, \dd f_k)
	\comm{\comm{\dots \comm{D}{\hat f_k}}{\dots\,,\hat f_2}}{\hat f_1}
\end{equation}
where, with a slight abuse of notation, we can identify a function with its multiplication operator (observe that all constants are in the center of the graded Lie algebra of differential operators). Due to the Jacobi identity, $\sigma(D)$ is symmetric. As a consequence, any second-order differential operator on $\mathcal M$ with a nondegenerate principal symbol gives rise in a natural way to a pseudo-Riemannian metric~\cite{vinogradov}.

\subsection{A result on Laplace--Beltrami operators}\label{sec:DaleVsLB}
In this section we prove Eq.~\eqref{eq:wavelaplace}, i.e.\ the fact that  the wave operator coincides with the (negative of the) Laplace--Beltrami operator when restricted on the exterior algebra of differential forms, in a setting slightly more general than the one in the main text.
Let $\mathcal{M}$ be a parallelizable manifold of dimension $m$ endowed with a global frame $(X_a )$ of \emph{commuting} vector fields, and let us denote by $\alpha^a $ the global coframe  characterized by
\begin{equation}\label{eq:dualitade}
	\inn[X_b ]\alpha^a =\ \alpha^a  (X_b )=\delta^a_b \,.
\end{equation}
By using the intrinsic expression of the exterior derivative of a 1-form given in Eq.~\eqref{eq:closed}, 
\begin{equation}
	\dd{\alpha^a }(X_b ,X_c )
	=X_b  \qty(\alpha^a  (X_c ))
	-X_c  \qty(\alpha^a  (X_b))
	-\alpha^a (\comm{X_b }{X_c })\,,
\end{equation}
we conclude that the basis 1-forms are closed,  hence by Eq.~\eqref{eq:dualitade} they also satisfy $\lie[X_b ]\alpha^{a }$. By Leibniz rule, these properties extend to any basis $p$-form $\alpha^{a_1\dots a_p}=\alpha^{a_1}\wedge\dots\wedge\alpha^{a_p}$:
\begin{align}
	\dd{\alpha^{a_1\dots a_p}}=0\,,\qquad\lie[X_b ] \alpha^{a_1\dots a_p}=0\,.
\end{align}
Now let $g^{a b }$ be a symmetric and non-singular square matrix, and let us denote by 
\begin{equation}
	L=g^{a b }\lie[X_a ]\lie[X_b ]\,,
\end{equation}
a liner differential operator on $\mathcal{M}$, which corresponds to the wave operator when $g^{ab}=\eta^{ab}$, see Eqs.~\eqref{eq:dalembert}--\eqref{eq:MinkT}. As explained in Sec.~\ref{sec:exterior}, we can now introduce a contravariant metric on $\mathcal M$ by using the principal symbol of $L$, 
\begin{align}\label{eq:controvaria}
	\tilde g=g^{a b }\, X_a \otimes X_b \,,
\end{align}
yielding the covariant metric $g =g_{a b }\,\alpha^a \otimes \alpha^b$,
where $g^{a c }g_{c b }=\delta^a_b $. Since the coefficients $g_{ab}$ in the above expression do not depend on the point, we can assume without loss of generality that $\abs{g_{ab}}=\delta_{ab}$, i.e.\ that the frame and the coframe are orthonormal. Then, we choose $\Omega_g=\alpha^{1\dots m}$ as the metric volume form. This allows us to consider the Hodge star operator on $\Omega(\mathcal{M})$, and to define the codifferential, see  Eq.~\eqref{eq:codiff}, and the Laplace--Beltrami operator $\dd\codd+\codd\dd$ determined by $g$.

The aim of this section is to show that the following identity  
\begin{equation}\label{eq:goal}
	\mathop{(\codd\dd +\dd\codd)}\omega =-L \omega\,,
\end{equation}
holds for all differential forms $\omega\in\Omega(\mathcal M)$. 
To this scope, we start by observing that, since the basis $p$-forms $\alpha^{a_1\dots a_p}$ with strictly-increasing indices make up a global frame of $\Omega^p(\mathcal M)$, by linearity it will be sufficient to show that the above equation holds for all forms of the kind $f\, \alpha^{a_1\dots a_p}$, with $f\in\mathcal{F}(\mathcal{M})$. From
\begin{equation}\label{eq:expadf}
	\dd f= \dd f (X_a )\, \alpha^a 
	=X_a (f)\,\alpha^a 
	=\lie[X_a ] (f) \,\alpha^a 
\end{equation}
and the closedness of the basis forms we immediately obtain that
\begin{equation}\label{eq:d}
	\dd \qty(f\alpha^{a_1\dots a_p})
	=\lie[X_c ] (f) \,\alpha^{c a_1\dots a_p}\,.
\end{equation}
Since $\alpha^a$ is orthonormal, if $(a_1,\dots, a_p,b_1,\dots, b_{m-p})$ is a permutation of $\qty{1,\ldots, m}$,  the Hodge dual of $ \alpha^{a_1\dots a_p}$ is a multiple of  $\alpha^{b_1\dots b_{m-p}}$, hence $\dd\hodge \alpha^{a_1\dots a_p}=0$ and $\codd\alpha^{a_1\dots a_p}=0$.
On the other hand, we obtain that
\begin{align}\label{eq:deltai}
	\codd \qty(f\alpha^{a_1\dots a_p})
	&=(-1)^{m(p+1)+s+1} \,\hodge \dd\hodge(f\alpha^{a_1\dots a_p})\\
	&=(-1)^{m(p+1)+s+1} \,\lie[X_c ] (f)\,
	\hodge\qty( \alpha^c  \wedge \hodge\alpha^{a_1\dots a_p})\\
	&=(-1)^{m(p+1)+s+1} \,\lie[X_c ] (f)\,
	(-1)^{m-p}\hodge\qty( \hodge\alpha^{a_1\dots a_p}\wedge\alpha^c  )\\
	&=(-1)^{(m+1)p+s+1} \,\lie[X_c ] (f)\,
	\hodge\qty( \hodge\alpha^{a_1\dots a_p}\wedge\alpha^c  )
	\\\label{eq:delta1}
	&=(-1)^{(m+1)p+s+1} \,\lie[X_c ] (f)\,
	\inn[(\alpha^c )^\sharp]\,\hodge\hodge\alpha^{a_1\dots a_p}
	\\\label{eq:delta2}
	&=(-1)^{(m+1)p+s+1} \,\lie[X_c ] (f)
	\,(-1)^{p(m-p)+s}\,\inn[(\alpha^c )^\sharp]\,\alpha^{a_1\dots a_p}\\
	\label{eq:deltaf}
	&=-\lie[X_c ] (f)\,\inn[(\alpha^c )^\sharp]\,\alpha^{a_1\dots a_p}\,,
\end{align}
where equality~\eqref{eq:delta1}  follows from Eq.~\eqref{eq:InuHodge}, whereas equality~\eqref{eq:delta2} is the duality property~\eqref{eq:HodgeDuality}.
Since in our settings $(\alpha^c )^\sharp=g^{c d }X_d $, we further obtain that
\begin{equation}
	\dd\inn[(\alpha^c )^\sharp]\,\alpha^{a_1\dots a_p}
	=g^{c d } \,\lie[X_d ]\alpha^{a_1\dots a_p}=0\,.
\end{equation}
Finally, by using Eqs.~\eqref{eq:d} and~\eqref{eq:deltaf}, as well as the Leibniz rule for the interior product and the Lie derivative, we get
\begin{align}
	\qty(\codd\dd +\dd\codd)\qty(f\alpha^{a_1\dots a_p})
	&=-\lie[X_c ]\qty(\lie[X_d ] \qty(f))
	\,\qty[\inn[(\alpha^d )^\sharp]\,\alpha^{c a_1\dots a_p}
	+\alpha^c \wedge \inn[(\alpha^d )^\sharp]\,\alpha^{a_1\dots a_p}]\\
	&=-\lie[X_c ]\qty(\lie[X_d ] \qty(f))
	\,\qty(\inn[(\alpha^d )^\sharp]\,\alpha^{c } ) \,\alpha^{a_1\dots a_p}\\
	&=-\lie[X_c ]\qty(\lie[X_d ] \qty(f))
	\,g^{d c } \,\alpha^{a_1\dots a_p}\\
	&=-g^{d c }\,\lie[X_c ] \lie[X_d ] \qty(f \alpha^{a_1\dots a_p})
	\,,
\end{align}
which is precisely Eq.~\eqref{eq:goal} for $\omega=f \alpha^{a_1\dots a_p}$.

\section{Decomposition of the exterior algebra}\label{app:projections}
In this appendix we describe the algebraic decomposition of $\Omega(\mathcal{M})$, where $\mathcal M$ is a generic differential manifold, with respect to a suitable pair of a 1-form $\xi$ and a vector field $X$.
We start by observing that the following identity holds:
\begin{equation}
	\xi(X) \,\omega=
	\inn[X] (\xi \wedge \omega) +\xi\wedge i_X\omega
	\,,
\end{equation}
where $\omega$ is an arbitrary differential form on $\mathcal M$. In other words, $\inn[X]  \est[\xi] +\est[\xi]\inn[X] $ is the multiplication by $\xi(X)$, where $\est[\xi]$ denotes the extension by the 1-form $\xi$, namely the linear map 
\begin{equation}
	\est[\xi](\omega)=\xi\wedge \omega\,,
\end{equation}
that is  homogeneous of degree 1 (i.e., maps $p$-forms to $(p+1)$-forms), and squares to zero.
From now on, we assume that $\xi$ and $X$ are nowhere-vanishing and contract to 1,
\begin{equation}\label{eq:Xxi}
	\xi (X)= 1\,.
\end{equation}
As a result, $\omega$ is decomposed as
\begin{equation}\label{eq:decoX}
	\omega
	=\xi\wedge\inn[X] \omega
	+\inn[X] \qty(\xi \wedge \omega)
	\,,
\end{equation}
that is, we have the resolution of the identity
\begin{equation}\label{eq:resolution}
	\Puno+\Pdue=\id_{\Omega(\mathcal M)}\,,
\end{equation}
where we have set
\begin{align}\label{eq:risolvo}
	\Puno=\est[\xi]\inn[X]\,,
	\qquad
	\Pdue=\inn[X]\est[\xi]\,.
\end{align}
The maps $\Puno$ and $\Pdue$ are linear, and homogeneous of degree 0. Since, for instance, $\inn[X]$ squares to zero, we have $\inn[X]\Pdue=0$, hence $\Puno \Pdue=0$. But then, since $\Puno$ and $\Pdue$ resolve the identity, they are also idempotent, and $\Pdue\Puno =0$, that is
\begin{align}\label{eq:comp}
	\Puno^2=\Puno\,,
	\qquad
	\Pdue^2=\Pdue\,,
	\qquad
	\Puno \Pdue=\Pdue \Puno=0\,.
\end{align}
In other words, given the resolution of the identity~\eqref{eq:resolution}, the above equalities are all equivalent to each other, and show that the image of each map is precisely its invariant subspace, that in turn is the kernel of the other map:
\begin{align}\label{eq:KerIm}
	\Im \Puno=\ker(\Puno-1)=\ker \Pdue\,,
	&&
	\Im \Pdue=\ker(\Pdue-1)=\ker \Puno\,,
\end{align}
Since, in particular, $\Im\Puno$ and $\Im\Pdue$ are linearly independent and span the entire exterior algebra, we have that
\begin{equation}
	\Omega(\mathcal M)=\Im \Puno \oplus \Im \Pdue\,,
\end{equation}
hence $\Puno$ and $\Pdue$ can be regarded as the algebraic projections of the above decomposition.

In the remaining part of this appendix, we will derive further properties of this decomposition.  In particular, we will discuss the behaviour of the wedge product, of the Lie derivative and of the Hodge star operator with respect to the decomposition brought about by $\Puno$ and $\Pdue$. 
Notice that the properties described in Sec.~\ref{app:compatibility}, in particular, require additional assumptions, namely that $\dd \xi=0$, that $\mathcal{M}$ is an orientable manifold endowed with a metric $g$ such that  $g(X,X)$ does never vanish and that $\xi=g(X,X)^{-1} X^\flat$. These assumptions are guaranteed to hold  in the main text of this paper, where we consider $X=\partial/\partial z$ and $\xi=\dd z$ or $X=\partial/\partial y$ and $\xi=\dd y$.
We refer the interested reader to \cite{Fec97, AuKu14, Koc16, CDML19} for further details.

\subsection{Algebraic properties}
Let $\omega$ and $\nu$ be any two differential forms. By noticing that 
\begin{equation}
	\Puno \omega\wedge\Puno \nu=0
\end{equation}
and by using the identity resolution in Eq.~\eqref{eq:risolvo}, we readily obtain the decomposition of their wedge product,
\begin{equation}\label{eq:PQwedge}
	\omega\wedge \nu
	=\Puno\omega\wedge \Pdue\nu +\Pdue\omega\wedge \Puno\nu +\Pdue\omega\wedge \Pdue\nu 
	\,,
\end{equation}
along with the identities
\begin{align}
	\Puno(\omega\wedge\nu)
	=\Puno\omega\wedge \Pdue\nu+ \Pdue\omega\wedge \Puno\nu\,, &&\Pdue(\omega\wedge\nu)
	=\Pdue\omega\wedge \Pdue\nu\,.
\end{align}
The above equations imply that  $\Puno$ is a derivation of degree zero, as opposed to $\Pdue$, which simply distributes with respect to the wedge product.    

Let us now consider two pairs of a 1-form and a vector field contracting to 1, that is $\xi^i(X_i)=1$  with $i=1,2$.
In general, operations of the same kind anticommute,
\begin{align}
	i_{X_i} i_{X_j}= i_{X_j\wedge X_i}=- i_{X_i\wedge X_j}=-i_{X_j} i_{X_i}\,,&&
	\epsilon_{\xi^i} \epsilon_{\xi^j} =\epsilon_{\xi^i\wedge \xi^j} =-\epsilon_{\xi^j\wedge \xi^i} =-\epsilon_{\xi^j}\epsilon_{\xi^i}\,,
\end{align}
the case $i=j$ reiterating the fact that the contraction by a vector field, and the extension by a 1-form, square to zero.  
On the other hand,
\begin{equation}
	i_{X_i }\epsilon_{\xi_j}
	= \xi^j\qty(X_i)\, \id_{\Omega(\mathcal M)}
	- \epsilon_{\xi^j}i_{X_i }\,,
\end{equation}
so that
\begin{equation}
	\mathcal P_i \mathcal P_j
	=\epsilon_{\xi^i} i_{X_i}\epsilon_{\xi^j} i_{X_j}
	= \xi^j\qty(X_i)\, \epsilon_{\xi^i}i_{X_j} 
	-\epsilon_{\xi^i}\epsilon_{\xi^j}i_{X_i} i_{X_j}
	\,,
\end{equation}
hence
\begin{equation}\label{eq:PiPj}
	\comm{\mathcal P_i }{\mathcal P_j}
	=\xi^j\qty(X_i)\,\epsilon_{\xi^i}i_{X_j} -
	\xi^i\qty(X_j)\,\epsilon_{\xi^j}i_{X_i}\,.
\end{equation}
In particular, if $\xi^j\qty(X_i) =\delta^j_i$, then  $\mathcal P_i$ and $\mathcal P_j$ commute.

\subsection{Lie derivative and Hodge star operator}\label{app:compatibility} 
From now on, we shall assume that $\xi$ is closed, $\dd\xi=0$. As an immediate consequence,
\begin{equation}\label{eq:Liexi}
	\lie[X]\xi=\dd \inn[X]\xi + \inn[X]\dd\xi=\dd 1=0\,.
\end{equation}
which in turn implies 
\begin{equation}\label{eq:PILie}
	\est[\xi] \lie[X]  \omega
	=\xi\wedge\lie[X] \omega
	=\lie[X] \qty(\xi\wedge\omega)
	=\lie[X]\est[\xi]  \omega \,,
\end{equation}
that is, $\lie[X]$ commutes with $\est[\xi]$. But then, since $\lie[X]$ always commutes with $\inn[X]$, it also commutes with both projections:
\begin{align}\label{eq:LieProj}
	\Puno \lie[X] =\lie[X] \Puno\,, \qquad \Pdue \lie[X] =\lie[X] \Pdue\,.
\end{align}
Therefore, in particular, $\lie[X]\omega=0$, if and only if $\lie[X]\Puno\omega=0$ and $\lie[X]\Pdue\omega=0$.

Let us further assume that $\mathcal M$  is an orientable manifold endowed with a metric $g$, having negative index of inertia $s$, and let $\Omega_g$ be the metric volume form. In the presence of a metric, we are allowed, so to say, to contract a differential form $\omega$ by another differential form $\nu$, by considering $i_{\nu^\sharp}\omega$. In these settings, we always have that $i_{\nu^\sharp}\hodge \omega=\hodge (\omega\wedge \nu)$, see Eq.~\eqref{eq:InuHodge}.
Moreover, if $\omega$ is a $p$-form, and $\nu$ is a $q$-form, we also have
\begin{align}
	\hodge \inn[\nu^\sharp] \omega 
	&= (-1)^{p(m-p)+s}\, \hodge i_{\nu^\sharp} \hodge\hodge\omega 
	\\
	&=(-1)^{p(m-p)+s}\, \hodge\hodge (\hodge \omega \wedge \nu)
	\\
	&=(-1)^{p(m-p)+s}\, (-1)^{(p -q)(m-p+q)+s}\, \hodge \omega \wedge \nu
	\\\label{eq:HodgeInu}
	&=(-1)^{q(p + q)}\,\nu  \wedge \hodge \omega\,,
\end{align}
where we made use of Eq.~\eqref{eq:InuHodge}, as well as of the duality property~\eqref{eq:HodgeDuality}.
We have just observed that, in the presence of a metric, the extension and the contraction are mutually intertwined (modulo a sign) by the Hodge star operator: if $\omega$ and $\nu$ are homogeneous of degree $p$ and $q$, respectively,
\begin{gather}\label{eq:InuHo}
	i_{\nu^\sharp}\hodge \omega= (-1)^{pq}\,\hodge \epsilon_\nu \omega\,,
	\\\label{eq:HoInu}
	\hodge  i_{\nu^\sharp}\omega 
	=(-1)^{q(p + q)}\,\epsilon_\nu \hodge \omega\,.
\end{gather}

Now, let us assume that $\xi$ and $X$ are related by $\xi^\sharp = f X$, with $f\in\mathcal F(\mathcal M)$. Then, by Eq.~\eqref{eq:Xxi}, we have $1= f\, g(X,X)$, therefore the function $g(X,X)$ must be nonvanishing (a constraint on $X$ involving the pseudo-Riemannian structure of the manifold), in which case there exists only one 1-form which is at the same time dual to $X$, in the sense of Eq.~\eqref{eq:Xxi}, and pointwise proportional to $X^\flat$:
\begin{equation}\label{eq:xiX}
	\xi=\frac{1}{g(X,X)} X^\flat\,.
\end{equation}
Hereafter, we shall assume that the ``norm'' $g(X,X)$ of $X$ is nowhere vanishing, and that the above equation holds.  Then, if $\omega$ is a $p$-form, Eqs.~\eqref{eq:InuHo}--\eqref{eq:HoInu} specialize to
\begin{gather}
	\inn[X]\hodge \omega= f^{-1}\,(-1)^{p}\,\hodge \epsilon_\xi \omega\,,
	\\
	\epsilon_\xi \hodge \omega 
	=f\,(-1)^{p + 1}\,\hodge  \inn[X]\omega \,.
\end{gather}
The above equations imply that $\inn[X]\hodge\est[\xi]=0$  and $\est[\xi] \hodge\inn[X]=0$,which in turn entail $\Puno\hodge \Puno=0$ and $\Pdue\hodge \Pdue=0$.
In other words, we have observed that
\begin{equation}\label{eq:offdiag}
	\hodge=\Puno\hodge \Pdue+\Pdue\hodge \Puno\,,
\end{equation}
which represents the decomposition of the Hodge star operator with respect to the projectors $\Puno$ and $\Pdue$.

\end{document}